\documentclass[floatfix, twocolumn,nofootinbib,longbibliography,preprintnumbers,10pt]{revtex4}

\usepackage{braket} 

\usepackage{graphicx}

\usepackage{amsmath}
\usepackage{amssymb}

\usepackage{enumerate}

\usepackage{chngcntr}

\counterwithin{paragraph}{section}

\newcommand{\numberfield}[1]{\ensuremath{\mathbb{#1}}} 

\newcommand{\vect}[1] {\ensuremath{\mathbf{#1}}} 

\newcommand{\id}{\ensuremath{\text{\tiny\textsf{ID}}}}   

\newcommand{\sid}{\ensuremath{_\id}}   

\newcommand{\psiid}{\ensuremath{\psi\sid}}

\newcommand{\tpsiid}{\ensuremath{\widetilde{\psi\sid}}}

\newcommand{\tableheadtext}[1]{\textsf{\textbf{\small #1}}} 
\newcommand{\tabletext}[1]{\small \textrm{#1}} 

\newcommand{\prop}[1]{\textsf{#1}} 

\newcommand{\Gmat}[4]{\ensuremath G(#1,#2, #3, #4) }

\begin{document}


\title{Persistence and Nonpersistence as Complementary Models \\ of Identical Quantum Particles}
\author{Philip Goyal}	
    \email{pgoyal@albany.edu}
    \affiliation{University at Albany~(SUNY), NY, USA}
\date{\today}
\begin{abstract}
According to our understanding of the everyday physical world, observable phenomena are underpinned by persistent objects that can be reidentified~(or tracked) across time by observation of their distinctive properties.  This understanding is reflected in classical mechanics, which posits that matter consists of persistent, reidentifiable particles. However, the mathematical symmetrization procedures used to describe identical particles within the quantum formalism have led to the widespread belief that identical quantum particles lack either persistence or reidentifiability.  However, it has proved difficult to reconcile these assertions with the fact that identical particles are routinely assumed to be reidentifiable in particular circumstances.  For example, when two electrons move through a bubble chamber, each is said to generate a sequence of bubbles~(a `track') that is caused by one and the same particle.   Moreover, neither of these assertions accounts for the mathematical form of the symmetrization procedures used to describe identical particles within the quantum framework, leaving open theoretical possibilities other than bosonic and fermionic behavior, such as paraparticles, which do not appear to be realized in nature.   Here we propose the novel idea that \emph{both} persistence and nonpersistence models must be employed in order to fully account for the behaviour of identical particles.  Thus, identical particles are neither persistent nor nonpersistent.  We prove the viability of this viewpoint by showing how Feynman's and Dirac's symmetrization procedures arise through a synthesis of a quantum treatment of these models, and by showing how reidentifiability emerges in a context-dependent manner.  We further show that the persistence and nonpersistence models satisfy the key characteristics of Bohr's concept of complementarity, and thereby propose that the behavior of identical particles is a manifestation of a \emph{persistence--nonpersistence complementarity,} analogous to Bohr's wave--particle complementarity for individual particles.  Finally, we construct a precise parallel between these two complementarities, and detail their conceptual similarities and dissimilarities.

\end{abstract}

\maketitle

We ordinarily conceive of the everyday physical world as consisting of \emph{objects} that  bear \emph{properties} and that \emph{persist} through time.  Developed early in life through our continual sensorimotor interaction with the physical world, this conception organizes our experience of the external world into a coherent, predictive model.  In particular, persistence underwrites our ability to say that the object one is seeing now is \emph{the same as} a specific object that one saw elsewhere at an earlier time. 
In practice, objects' gradual motion and slowly-varying characteristic properties~(such as shape and colour) provide the perceptual handles that enable their reidentification.

Classical physics incorporates these key notions---objects, properties, and persistence---into its abstract conceptual framework at a fundamental level.  Persistence is reflected in the assumption that objects can be \emph{labelled}.  Additionally, classical mechanics posits that objects localized to point-like regions of space---\emph{particles}---are the fundamental constituents of matter.    Much as in everyday experience, these particles can be reidentified by continuous tracking of their motion, and by measurement of their characteristic intrinsic properties~(such as mass and charge).  In this framework, two particles may be entirely \emph{identical} in their intrinsic properties---a situation that does not arise in everyday experience---yet remain reidentifiable by their distinct trajectories.  

It is, however, widely accepted that quantum theory challenges the validity of the classical particle-based understanding of the physical world.  
In particular, the quantum treatment of assemblies of identical particles has brought into question the assumptions of persistence and reidentifiability.  This challenge was first brought to light through Bose's derivation of Planck's blackbody radiation formula~\cite{Bose1924}.  In this derivation, calculation of the number of ways in which a given number of photons can be arranged amongst cells in phase space only takes into account the \emph{number} of photons in each cell.  Thus, unlike Boltzmann's corresponding calculation for gas molecules, no account is taken of \emph{which} photon is in which cell.

Bose's counting procedure admits two quite distinct interpretations.  First, that the photons are \emph{not persistent,} so that the very notion of `which photon is in which cell' is meaningless.  Second, that the photons are \emph{not reidentifiable} by any observer.  The first view was taken by many contemporary physicists.  For example, at the 1927 Solvay Conference, Langevin suggested that the novel quantum statistics pointed to a suppression of the `individuality of the constituents of the system'~\cite[p.~453]{BacciagaluppiValentini2009}.  More pointedly, in his 1950 Dublin lectures~\cite{Schroedinger1952}, Schroedinger states:~``If I observe a particle here and now, and observe a similar one a moment later at a place very near the former place, not only cannot I be sure whether it is `the same', but this statement has no absolute meaning.''   The second view is based on the \emph{symmetrization procedure} that was put forward by Heisenberg and Dirac~\cite{Heisenberg26, Dirac1926, Dirac30} as a way of incorporating Bose's novel counting procedure and Pauli's exclusion principle into the nascent quantum formalism.   According to Dirac~\cite[\S54]{Dirac58}, what is special about identical particles is that they are `indistinguishable'---or not reidentifiable---in the sense that observations provide no information about which particle is which.

However, these interpretations are both at odds with assumptions that are routinely made in the interpretation of primary experimental data.  For example, in an experiment in which we say that electrons are created at a filament, diffracted through a crystal lattice, and then impact a phosphorescent screen, we presume that each scintillation on the screen is due to the same electron that was emitted by the filament, \emph{even though there are many other electrons in the laboratory and elsewhere.}   The correctness of the diffraction pattern calculated on the assumption of persistence demonstrates that the assumption is at least \emph{approximately} validity in this instance.  Yet, according to Dirac's symmetrization procedure, this electron is in a symmetrized state with all the other electrons~(irrespective of its ostensible isolation from them), which implies that each has the same reduced state, so that each is equally likely to be found in inside or outside the experiment~(see, for instance,~\cite{Mirman73, Muynck1975, DieksLubberdink2011, Jantzen2011, Caulton2013}).  Similarly, the notion of `particle tracks' (say, in a bubble chamber), which is a prerequisite to the processing of primary data in particle physical experiments, implicitly assumes object persistence---a sequence of bubbles is deemed to have been generated by the \emph{same} particle, thus constituting a `track'---even when another particle identical to it lies simultaneously in the detector's field of view.  

Additionally, neither of these interpretations provide a basis for accounting in detail for the quantum rules employed in the treatment of identical particles.  For example, although the nonpersistence view naturally accounts for Bose's photon-counting procedure, it provides no clue as to the origin of Pauli's exclusion principle, which~(in a modification of Bose's procedure) was implemented by Fermi as a single-occupancy limit on each phase-space cell~\cite{Fermi1926}.  Similarly, although Dirac's non-reidentifiability view explains why a system initially placed in a symmetric or antisymmetric state will remain in the same type of state under temporal evolution, it provides no explanation of why a system cannot be in a nonsymmetric state~(specifically, in a linear combinations of symmetric and antisymmetric states) in the first place.  Dirac's view also leaves open the theoretical possibility that a system of three or more particles could exhibit so-called parastatistical behaviour, a possibility for which no experimental evidence has been found~(see~\cite[\S54]{Dirac58} and~\cite{MG64}).

The above considerations suggest that neither interpretation is satisfactory, and that a more thoroughgoing revision of our conceptual picture is necessary if we are to pinpoint the essential idea that underlies the behaviour of identical particles and rigorously account for the empirical success of Dirac's symmetrization procedure.  Now, as we have noted, identical particles \emph{sometimes} behave as if persistent~(for example, two electrons moving along distinct particle `tracks'), and sometimes as if nonpersistent~(as in Bose's photon-counting procedure).  This suggests that, rather than trying to account for this behavioral diversity on the basis of persistence \emph{or} nonpersistence alone, we instead attempt to combine \emph{both} of these pictures in a more even-handed way. 
 
In this paper, we develop a novel understanding of identical particles along these lines.  We adopt an operational approach in which the raw data consists of \emph{identical localized events}.  To be concrete, one can think of observing a fixed number of localized light-flashes of the same colour at successive times.  At this stage, there are no `identical particles' as such, only identical \emph{events.}  We construct two distinct models of these events, namely a \emph{persistence} model and a \emph{nonpersistence} model.  These differ in whether or not it is assumed that successive events are generated by individual persistent entities~(`particles').   We then show that these models can each be described within the Feynman formulation of quantum theory and be \emph{synthesized} to derive the Feynman's form of the symmetrization procedure~\cite{FeynmanHibbs65}.   As we show elsewhere~\cite{Goyal2015, Goyal2018b} and summarize here, this procedure can be transformed into a state-based symmetrization procedure which is empirically adequate yet differs from Dirac's procedure in form and meaning, in particular allowing for the natural emergence of reidentifiability in special cases.  

We then show that the persistence and nonpersistence models, and the manner of their synthesis, satisfy the key characteristics of Bohr's concept of complementarity, specifically that these models are mutually exclusive, and that they can be synthesized but not unified.  On this basis, we propose that the quantal behaviour of identical particles reflects a \emph{complementarity of persistence and nonpersistence,} analogous to the  way in which the behavior of an individual electron reflects wave--particle complementarity.  

Finally, we exhibit a precise parallel between our proposed persistence--nonpersistence complementarity and Bohr's wave--particle complementarity.  In particular, we show that the Feynman amplitude sum rule can be viewed as a synthesis of the wave and particle models of elementary constituents of matter in a manner that formally mirrors the way in which a symmetrization procedure arises through a synthesis of persistence and nonpersistence models of identical localized events.    These two examples thereby illustrate how complementarity can be turned into a precise methodology for synthesizing mutually-exclusive models.

The remainder of this paper is organized as follows.  In Sec.~\ref{sec:persistence}, we define the concepts of persistence and nonpersistence in an operational manner.  In Sec.~\ref{sec:reconstruction}, we outline the derivation of the symmetrization procedure.  In Sec.~\ref{sec:interpretation}, we describe our complementarity interpretation of identical particles in light of this derivation, and establish a detailed parallel with Bohr's wave--particle complementarity.   We conclude in Sec.~\ref{sec:discussion} with a discussion of the broader context and some open questions.


\section{Operational Framework}
\label{sec:persistence}

Our discussion will be based on the fundamental key notions of \emph{persistence} and \emph{nonpersistence}.  In order to place these notions on a clear footing, we begin by stepping back from the familiar theoretical frameworks of classical and quantum mechanics, and instead adopt an operational perspective.  Such a perspective is helpful in identifying assumptions that are of limited validity in physical domains remote from everyday experience, but that are too entrenched in our customary patterns of thought to be clearly and consistently perceived.  

\begin{figure}
\begin{center}
\includegraphics[width=3in]{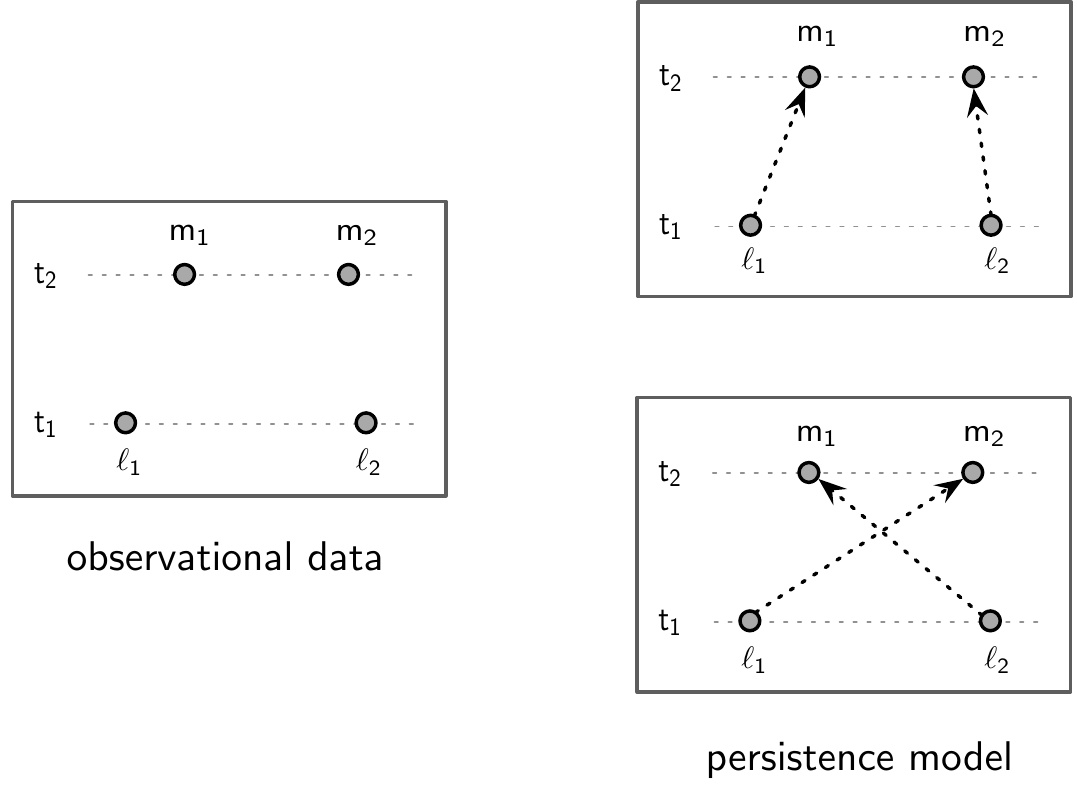}
\caption{\label{fig:persistence-model}\emph{Event detections and the persistence model.}  \emph{Left:} Two localized events are detected at the successive times.   The measurements at time~$t_1$ and~$t_2$ yield outcomes~$\ell_1, \ell_2$ and~$m_1, m_2$, respectively, with the labelling convention that the left-most locations are~$\ell_1$ and~$m_1$.  An additional property---`colour' in our example---of each event is measured, according to which the events are said to be nonidentical or identical.  In the example here, the events are identical, illustrated by the two filled circles at each time.  The set of these additional properties at each time is conserved, illustrated here by there being two filled circles at each time.  \emph{Right:}  A \emph{persistence} model of the data posits that two persistent entities~(`particles') are responsible for these detections.  Each particle is ascribed colour as an intrinsic property, whose value is constant.  In the example here, the particles are the same colour, and are thus identical.  On this model, one can say either that the same particle is responsible for detections~$\ell_1$ and~$m_1$~(proposition~$\prop{A}$) or that the same particle is responsible for~$\ell_1$ and~$m_2$~(proposition~$\prop{B}$).}
\end{center}
\end{figure}

\paragraph{Position measurements of localized events and their properties.}  Consider a situation where a position measurement, implemented by a fine grid of detectors that tile a region of space, is performed at discrete times~$t_1, t_2,\dots$~(see Fig.~\ref{fig:persistence-model}).  Suppose that only two detectors fire at each time.  We can speak of each such detection as a \emph{localized event}~(a `flash').   Suppose further that these detectors are capable not only of registering a localized event, but also of measuring some additional properties of this event.  For concreteness, we henceforth imagine that there is just one additional property, namely \emph{colour;} thus, at each time, one observes \emph{two coloured flashes}.

Let us further suppose that observation shows that these additional properties are conserved, in the sense that the total number of flashes of each colour seen at each time is the same.  For example, at each time, one obtains a blue flash and a red flash.  If it should be the case that both of the localized events have the same additional property values~(for example, both the flashes are blue), we shall say they are \emph{identical}.  That is the situation that concerns us here.  Finally, let us suppose that the system is isolated.  Operationally, this can be established by carrying out repeated trials, and showing that the probability over the locations of the two detections at~$t_2$ are determined by the locations at~$t_1$, a condition we refer to as \emph{closure}~\cite{GKS-PRA}.

On the basis of these observations, we say that the measurements are being performed on a `system', namely a persistent object that is such that (i)~it yields two localized events at each time~$t_i$, and (ii)~the property-values~(colours) of these events is conserved. 

\subsection{Persistence model}

We now construct a model of data in which two identical localized events are registered at each time.  Let us assume that there exist individual entities that persist in between these detections, and that these entities can (informally) be said to \emph{cause} these detections.   
We shall refer to these entities as \emph{particles} on the understanding that this word describes the localized, particle-like way that they manifest themselves upon detection, rather than implying anything about their nature \emph{between} such detections.   We ascribe \emph{intrinsic} and \emph{extrinsic} properties to these particles.  The former are the same as the additional properties measured of the localized events, and their values are assumed to be constant.  In this case, the flashes are the same colour, so we say that there are two particles of the same colour, and we assume that the colour of each particle is constant.  As the colours of two particles are the same, we shall say that they are \emph{identical}.  

On the basis of this \emph{persistence model}, one can now meaningfully say that \emph{same} particle is responsible for detections at two different times.  Suppose that events are detected at locations~$\ell_1$ and~$\ell_2$ at time~$t_1$, and at~$m_1$ and~$m_2$ at~$t_2$.  Here, we adopt the labelling convention that~$\ell_1, m_1$ are the left-most locations at each time.   
According to the persistence model, even though the two events at each time are identical, one can say that one or the other of the following propositions is true:
\begin{quote}
\textsf{A = `the same particle is responsible for the detection~$\ell_1$ at~$t_1$ and the detection at~$m_1$ at~$t_2$,'}
\textsf{\emph{or}}

\textsf{B = `the same particle is responsible for the detection~$\ell_1$ at~$t_1$ and the detection at~$m_2$ at~$t_2$.'}  
\end{quote}

\begin{figure}
\begin{center}
\includegraphics[width=3in]{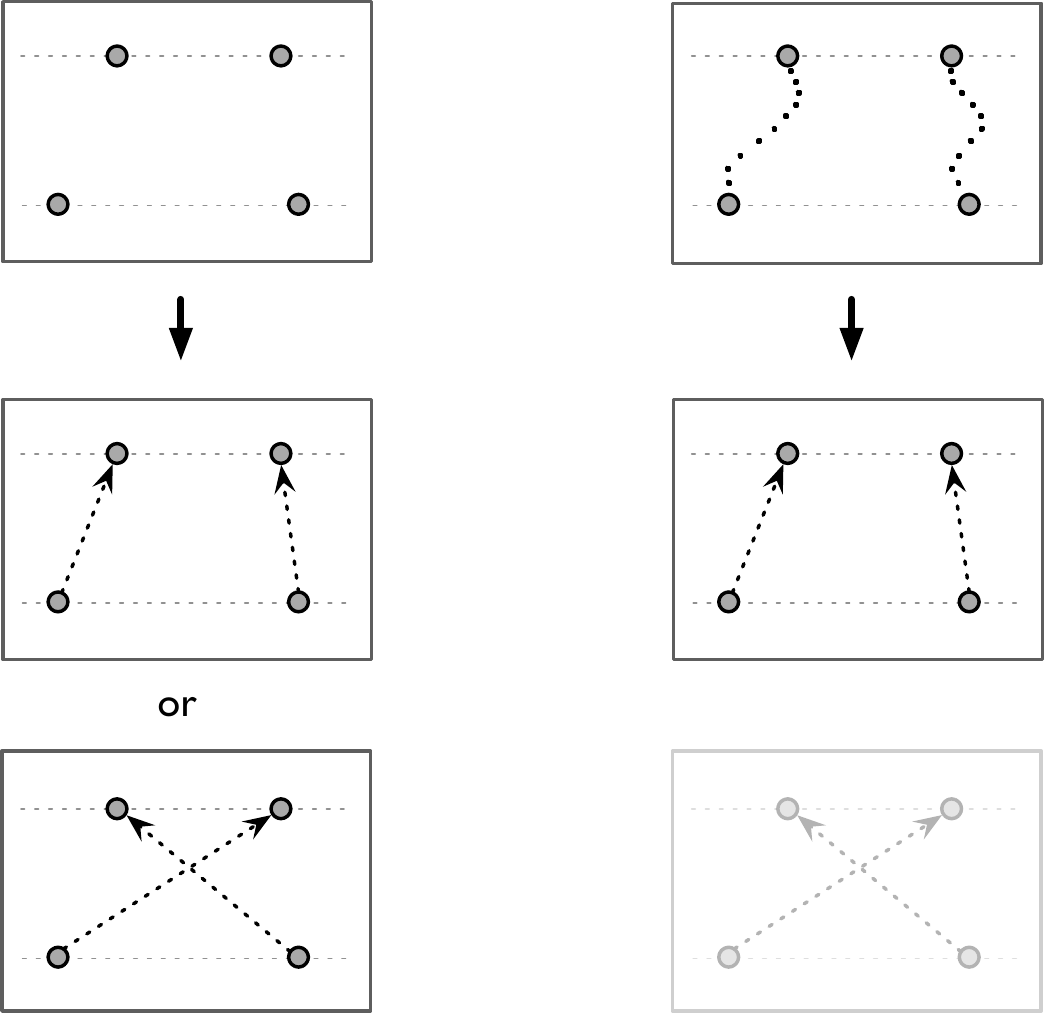}
\caption{\label{fig:reidentification}\emph{Reidentification of identical particles.}  Two identical localized events are detected at successive times.  A \emph{persistence} model of the data posits that two identical persistent entities~(`particles') are responsible for these detections.  Reidentification may or may not be possible, depending upon whether events at each time are identical, and on whether additional assumptions can be made about the nature of the particles.   \emph{Left:} As the events are identical~(same colours), reidentification is not possible on the basis of measurement of their intrinsic properties~(colours) given only observations at~$t_1$ and~$t_2$.  \emph{Right:}  If one has numerous observations between times~$t_1$ and~$t_2$, and if further modelling assumptions are made~(that the particles move continuously and are undisturbed by measurement), then approximate reidentification is possible.  In the idealized limit of an observer capable of observation of arbitrarily precision at arbitrarily high frequency~(as presumed in classical mechanics), perfect reidentification is possible.}
\end{center}
\end{figure}

\paragraph{Reidentifiability.}  If it is possible for an \emph{observer} to determine which of the above propositions is true, we shall say that it is possible to \emph{reidentify} each particle~(see Fig.~\ref{fig:reidentification}).  That is, reidentification is the observational counterpart of the theoretical notion of persistence.   As the particles are identical~(same colour), reidentification on the basis of measurement of their intrinsic properties is impossible, so that the possibility of reidentification will hinge upon additional theoretical and operational considerations.

In particular, if one makes the additional theoretical assumption that particle motion is continuous, and further assume that it is possible for an observer to make non-disturbing position measurements of arbitrary precision at arbitrarily high frequency, then reidentification~(at arbitrarily high confidence level) of identical particles is possible on the basis of the measurement records.

\paragraph{Persistence grounds particle labels.}
If persistence is assumed, then either proposition~\prop{A} or~\prop{B}, as given above, is true.  This provides the basis for particle labelling.  Specifically, let us label~`1' the particle that was at~$\ell_1$ at~$t_1$, and label the other particle~`2'.  Then, the \emph{configuration} of the system at~$t_1$ is given by the ordered pair~$(\ell_1, \ell_2)$, and the configuration at time~$t_2$ is either~$(m_1, m_2)$ or~$(m_2, m_1)$.  Each of the latter two ordered pairs thus reflects not only the observed particle positions at~$t_2$, but also---as a result of the theoretical assumption of persistence---\emph{some} information about the observed positions at the earlier (reference)~time,~$t_1$.  For example, the configuration~$(m_1, m_2)$ specifies not only the information that there are two identical particles at locations~$m_1$ and~$m_2$ (which is what is observed at~$t_2$), but also that the particle that is now at~$m_1$ was earlier at~$\ell_1$. 

To say that reidentification is, in principle, possible, means that there exists an observer who can determine which of these configurations is in fact the case.  But, for an observer who is not capable of reidentification, there is a gap between the theoretical level of description of the system---the configuration---and the information available to that observer.  For example, at~$t_2$, the theoretical description might be the configuration~$(m_1, m_2)$, but the observer would be incapable of distinguishing this from~$(m_2, m_1)$.

\paragraph{Connection to Classical Mechanics.}
In classical mechanics, persistence is assumed, and an ideal observer is capable of reidentifying identical particles~(provided they cannot coincide) by tracking their continuous trajectories with arbitrary precision but without causing disturbance.  
Therefore, the theoretical and observational descriptions coincide in the configuration~$(\vect{r}_1, \vect{r}_2)$ of two particles.

\subsection{Nonpersistence Model}

We can, however, construct a second model---a \emph{nonpersistence model}---of the identical localized events in which one does \emph{not} presume that there are individual persistent entities that underlie the individual localized detections.   Rather, the two localized events at each time are regarded as a manifestation of a single abstract `system' that persists.   One is thus left with the bare data of localized events~$\{\ell_1, \ell_2\}$ at~$t_1$ and~$\{m_1, m_2\}$ at~$t_2$.  Repeated trials of the experiment would yield a conditional probability distribution~$\Pr\left( \{m_1, m_2\} \,|\, \{\ell_1, \ell_2\} \right)$.  

In such a model, the only persistent object is an abstract `system' which yields two localized detection events at each measurement time.  As this object is not analyzed into two separate persistent objects, the fundamental techniques that one ordinarily employs in constructing particle-based models are unavailable.  For instance, in classical mechanics, one  can start by positing that individual objects move uniformly if isolated, and then build up a model of a system of how two such objects interact with one another by imposing constraints in the form of conservation laws; but such a model-building strategy hinges on an analysis of the system into persistent individuals which is unavailable in the nonpersistence model.

\section{Derivation of a Symmetrization Procedure}
\label{sec:reconstruction}

As indicated in the Introduction, the quantum treatment of identical particles brings the assumption of persistence into question.  If one were to refrain from making this assumption, one would be left with the observational data, which for identical localized events consists simply of a list of the event-positions at each time.  But, on the basis of such data alone, it is not clear how one could go about generating predictions about the event-positions at a later time since a particle-based model-making strategy is blocked.

On the other hand, we have observational evidence that persistence is at least approximately valid in certain situations, for example in the case of identical localized events in a bubble chamber.  In order to  construct a predictive model, we incorporate \emph{both} of these pieces of observational evidence by formulating \emph{two} models of these events---one that assumes non-persistence and the other that assumes persistence---and then connect them.

\subsection{Synthesis of Persistence and Nonpersistence}

\begin{figure}
\begin{center}
\includegraphics[width=3.5in]{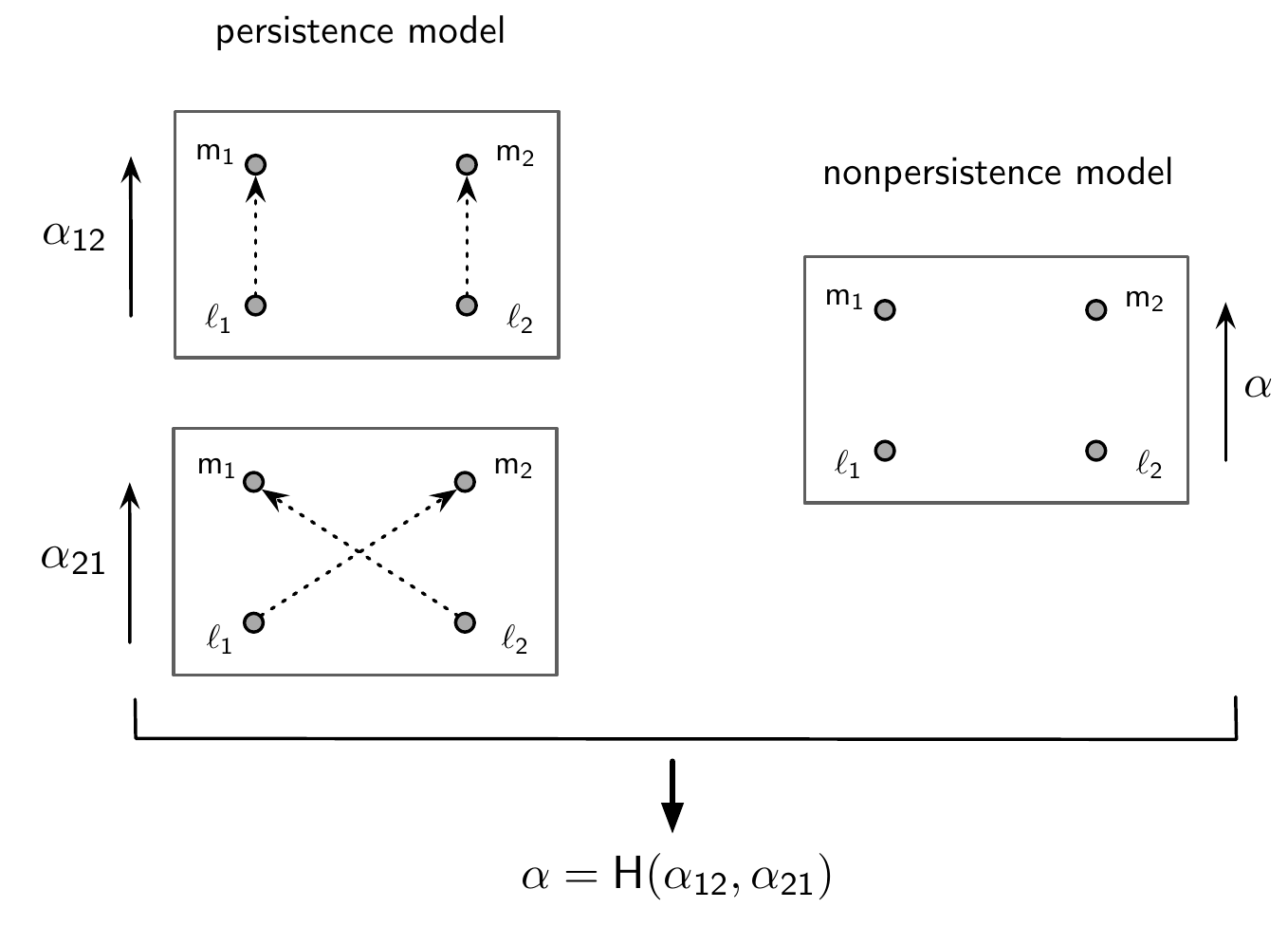}
\caption{\label{fig:SP-feynman}\emph{Derivation of Feynman's symmetrization procedure for two identical localized events.}  Position measurements at times~$t_1$ and~$t_2$ yield outcomes~$\ell_1, \ell_2$ and~$m_1, m_2$, respectively.   Two models---a persistence model and a nonpersistence model---of this data are constructed.   (i)~\emph{Left:}~According to the persistence model, persistent `particles' are responsible for the individual detections.  The figures on the left show the transitions of two identical particles compatible with these outcomes according to this  model:~the `direct' transition of amplitude~$\alpha_{12}$, and the `indirect' transition of amplitude~$\alpha_{21}$.   (ii)~\emph{Right:}~In the nonpersistence model, it is meaningless to say that a given detection at~$t_2$ was caused by the same thing as given detection at~$t_1$.  Accordingly, the figure on the right shows the only transition amplitude,~$\alpha$, that one can meaningfully assign according to this model.  The \emph{operational indistinguishability postulate}~(OIP) posits the relation~$\alpha = H(\alpha_{12}, \alpha_{21})$ between the amplitudes in these models, where~$H$ is a complex-valued function to be determined.  In Ref.~\cite{Goyal2015}, it is shown that~$\alpha = \alpha_{12} \pm \alpha_{21}$, with the sign corresponding to bosonic or fermionic behaviour.}
\end{center}
\end{figure}

To be specific, consider again the above experiment involving position measurements at times~$t_1$ and~$t_2$.   We now construct two theoretical models of this situation within the Feynman quantum formalism~(see Fig.~\ref{fig:SP-feynman}).  In the \emph{persistence} model, irrespective of whether or not reidentification is possible, we can say that either one or the other transitions connects the observed data---a \emph{direct} transition, where the particle that was at~$\ell_1$ at~$t_1$ is later found at~$m_1$ at~$t_2$; or an \emph{indirect} transition from~$\ell_1$ to~$m_2$.  Let us denote the amplitudes of these transitions~$\alpha_{12}$ and~$\alpha_{21}$, respectively.

The second model, the \emph{nonpersistence} model, does \emph{not} presume that there are persistent entities that underlie the observed localized detections.   Accordingly, the only amplitude that one can associate with the given data in this model is the transition amplitude,~$\alpha$, from the initial data~$\{\ell_1, \ell_2\}$ at~$t_1$ to the final data~$\{m_1, m_2\}$ at~$t_2$.
The connection between these two models takes the form of the \emph{operational indistinguishability postulate}~(OIP), which posits that the amplitudes in the persistence model determine the amplitude in the nonpersistence model.  In the case of measurements at two successive times under consideration here,
\begin{equation} \label{eqn:OIP}
\alpha = H(\alpha_{12}, \alpha_{21}),
\end{equation}
where~$H$ is an unknown continuous complex-valued function to be determined. 

The OIP also applies to the case where one has observations at \emph{three} successive times,~$t_1, t_2$ and~$t_3$.  In that case, the persistence model has two possible transitions between times~$t_1$ and~$t_2$, with amplitudes~$\gamma_{11}, \gamma_{12}$, and two possible transitions between times~$t_2$ and~$t_3$, with amplitudes~$\gamma_{21},$ and~$\gamma_{22}$.  The generalization of Eq.~\eqref{eqn:OIP} thus reads
\begin{equation} 
\label{eqn:G-def}
\gamma = \Gmat{\gamma_{11}}{\gamma_{12}}{\gamma_{21}}{\gamma_{22}},
\end{equation}
where~$\gamma$ is the transition amplitude in the nonpersistence model, and~$G$ is a function to be determined.  

We now incorporate the fact that there are situations in which we commonly say that the particle observed now is the `same' as one previously observed via an \emph{isolation condition}.  
This condition stipulates that, in the limiting case that isolation obtain for one of more of the identical particles in a given system, they can be treated as a \emph{persistent subsystem} for the purpose of making predictions.  For example, if the electron in a hydrogen atom is effectively isolated from all other electrons in a given system, then we can infer that \emph{the same} electron is responsible for the successive electron-detections in the atom.  One can accordingly apply the quantum formalism to this electron as if it were a (persistent)~system.   Formally, for the case of two identical events, the isolation condition requires that the transition probability~$|H(\alpha_{12}, \alpha_{21})|^2$ is the same as the probability of the persistence-model transition that has non-zero probability.  For example, if the direct transition is the one with non-zero probability, then 
\begin{equation}
|H(\alpha_{12}, 0)|^2 = |\alpha_{12}|^2.  
\end{equation}

From the assumptions above, Feynman's symmetrization procedure can be derived~\cite{Goyal2015}.   The key idea behind the derivation is the recognition that the amplitude for a particular process in the nonpersistence model can sometimes be computed in two different ways, and, in these instances, consistency of the assumptions implies that these calculational paths must agree.  Each such call for consistency leads to a functional equation.  For example, one obtains
\begin{multline} \label{eqn:G-product}
\Gmat{\alpha_{12} \beta_{12}}{\alpha_{12}\beta_{21}}{\alpha_{21}\beta_{12}} {\alpha_{21}\beta_{21}}  = \\ H (\alpha_{12}, \alpha_{21}) \, H(\beta_{12}, \beta_{21}).
\end{multline}
Solution of these functional equations yields
\begin{equation} \label{eqn:SP}
\alpha = \alpha_{12} \pm \alpha_{21},
\end{equation}
where the~$\pm$~sign corresponds to bosonic or fermionic behavior.  This is Feynman's symmetrization rule for two particles.  The above derivation generalizes naturally to~$N$ identical particles. 

\subsection{Probabilistic Reidentifiability} \label{sec:probabilistic-reidentifiability}  If one of the transition probabilities~$|\alpha_{12}|^2$ or~$|\alpha_{21}|^2$ is much smaller than the other, then the transition probability~$|H(\alpha_{12}, \alpha_{21})|^2$ approximates to the largest of these probabilities.  In that case, one can treat the observational data as a probabilistic version of the persistence model, so that one has \emph{probabilistic reidentifiability} of the particles.  Such a situation obtains, for example, for two electrons in the field of view of a particle chamber, where we can roughly say that the each electron travels along its own `track', even though there is a finite probability~(as computed using the persistence model) of the electrons `swapping' between tracks.

\subsection{State representation of the symmetrization procedure}

The amplitude-based symmetrization procedure given above can be re-expressed in terms of states and observables, a more familiar arena for the description of quantum phenomena.  Such a re-expression also facilitates a direct comparison with Dirac's symmetrization procedure.  Here we summarize the main ideas, referring the reader to~\cite{Goyal2015, Goyal2018b} for details.  To illustrate the key ideas, it suffices to consider two particles moving in one dimension.  In that case, one can re-express Eq.~\eqref{eqn:OIP} in terms of states as
\begin{equation} \label{eqn:SP-state}
\psiid(x_1, x_2) = \psi(x_1, x_2) \pm \psi(x_2, x_1) \quad\quad x_1 \leq x_2.
\end{equation}
The function~$\psiid$ on the left-hand side is the state in the nonpersistence model.  It is defined over location-space~(or reduced configuration space), namely~$x_1 \leq x_2$, and is normalized over that space.   The function~$\psi$ on the right-hand side is a state in the persistence model, and is defined over all~$\numberfield{R}^2$.  The above relation is to be understood as a \emph{connection} between these two models---it is a means of taking states from the persistence model over to the nonpersistence model.  Consequently, this relation requires careful reading as~$x_i$ has a \emph{different physical meaning} on each side of the equation.  On the right, the first argument of~$\psi$ is the $x$-location of particle~$1$, and the second that of particle~$2$.  However, on the left,~$x_1$ is the location of the leftmost particle-like detection, \emph{not} the $x$-location of particle~$1$.  

To determine the evolution of the state in the nonpersistence model, one evolves the corresponding state in the persistence model using a Hamiltonian that is symmetric in the particle labels, and then applies the above relation.  The symmetry of the Hamiltonian reflects the fact that energy does not depend on how identical particles are labelled in the persistence model.  Operators corresponding to observable quantities are defined in the nonpersistence model.  For example, the operator~$x_1$ denotes the $x$-value of the left-most particle, whilst~$(x_2 - x_1)$ denotes the inter-particle distance.  Note that there is no restriction to symmetric observables in the nonpersistence model.  

If isolation obtains, then the above observables gain additional meaning.  For example, if the two particles~(as viewed in the persistence model) are confined to disjoint regions on the left and right sides, then~$\alpha$ reduces to~$\alpha_{12}$.  Further, owing to the isolation between the particles, the surviving amplitude can be written as a product of two amplitudes, one related to each particle.  Thus, in the state formulation,~$\psiid(x_1, x_2) = \phi_\text{a}(x_1) \phi_\text{b}(x_2)$, where~$\phi_\text{a}, \phi_\text{b}$ can be viewed as one-particle states of labelled particles which have no common support.  Under these circumstances, the observable~$x_1$ in the nonpersistence model, which by default is the $x$-location of the left-most particle, gains the \emph{additional} meaning of the $x$-location of particle~$1$. Thus, in this limiting case, one recovers reidentifiability, and one has justification to model each of the particles as a distinct entity without regard for the other.  Note that such a recovery of reidentifiability is not possible within Dirac's symmetrization procedure---two identical particles remain in a symmetrized state irrespective of their ostensible isolation, which implies that they have the same reduced state and that each is equally likely to be found in one location or the other~(see, for instance,~\cite{DieksLubberdink2011, Jantzen2011, Caulton2013}).

For convenience, one can formally extend~$\psiid(x_1, x_2)$ to the entire configuration space, to obtain state~$\tpsiid(x_1, x_2)$, in terms of which one can rewrite the above equation as
\begin{equation} \label{eqn:SP-state'}
\tilde{\psiid}(x_1, x_2) =  \frac{1}{\sqrt{2}} \bigl[ \psi(x_1, x_2) \pm \psi(x_2, x_1) \bigr],
\end{equation}
where now~$x_1, x_2$ range over~$\numberfield{R}^2$.  Such a formal extension is useful in the sense that now~$\tilde{\psiid}$ and~$\psi$ are both defined over~$\numberfield{R}^2$ and both formally live in the same space~(a tensor product of two labelled copies of a one-particle Hilbert space).  However, the formal extension makes the reading of the labels more complicated:~although~$x_1, x_2$ on the right hand side are particle labels as before, $x_1$ on the left is the location of the leftmost particle whenever~$x_1 \leq x_2$, but the location of the rightmost whenever~$x_1 > x_2$.  
A relation resembling the one above is commonly used in the practical application of Dirac's symmetrization procedure in order to generate a symmetrized state from a product state.  However, in that case, the product and symmetrized states are both interpreted in the same model.

\section{Complementarity of Persistence and Nonpersistence}
\label{sec:interpretation}

As formulated by Bohr, \emph{complementarity} expresses a view about the kind of theoretical models that one can formulate about microphysical physical phenomena~\cite{Bohr-complementarity, Feyerabend-complementarity}.  Its key features are as follows.
\begin{enumerate}
\item \emph{Need for two incompatible models.}  The adequate description of a particular domain of physical phenomena requires \emph{two} distinct models.  These models are mutually incompatible in the sense that one cannot consistently apply all of the concepts of both to a given physical situation.
\item \emph{Synthesis of models.}  These two models can be synthesized through the addition of new ideas and assumptions, with the synthetic model describing behaviour that in some sense interpolates between that permitted by the original models.  
\item \emph{Non-unification.}  The concepts of at least one of the original models form an integral part of the synthetic model.  Hence, the original models are synthesized, but not unified into a new conceptual structure.
\end{enumerate}

\subsection{Wave \& Particle}
We illustrate the three key features of complementarity via Bohr's paradigmatic example of the wave and particle models of the electron.

First, according to the particle model, the electron is a point-like entity that has a definite position at each time.  In contrast, the wave model treats the electron as a delocalized wave-like object\footnote{In classical physics, one typically regards a `wave'~(such as a water wave, or a wave on a string) as underpinned by local disturbances over a region of space.  In such cases, `wave' is merely a collective noun that refers to a set of synchronized local disturbances, and so has no fundamental existence in and of itself.  However, in the wave model of an electron, such an underpinning is \emph{not} presumed, so that `wave' is taken to refer to an unanalyzed, delocalized object.}.  Each of these models paints a clear, visualizable conception of the electron, but are manifestly incompatible---an electron cannot simultaneously be a localized point-like object and a delocalized object.  Yet, the need for both models is made plausible by experimental phenomena~(such as diffraction) associated with electrons.  

Second, the synthesis of these models was first proposed by de Broglie in his composite wave-particle model of the electron, with Schroedinger's wave mechanics its successful culmination.  The latter synthesis introduces the novel concept of a wavefunction, which allows behaviour that is, in some sense, \emph{intermediate} between wave and particle---a wave-packet may exhibit particle-like or wave-like behaviour, depending upon the size of the aperture through which it passes.  

Third, the synthetic model does not unify the particle and wave models since it makes essential use of the particle model, in three distinct senses.  First, the wave equation is obtained via a quantization of the classical particle model of the system.  Second, the notion of a \emph{position measurement}~(in the Born rule) is a reference to the particle model.  Its abstraction---a `measurement outcome'---in the Dirac--von Neumann quantum formalism similarly corresponds to the notion of particle localization in space and time.   Third, the particle model is essential to the demarcation of the phenomenon to which we refer as `an experiment', and to the interpretation of experimental data.  For example, in an experiment where we say that electrons are emitted by a filament, diffracted through a crystal, and detected as scintillations on a screen, the emission and detection events are localized events that are interpreted within the particle model as `measurements of position'.  These events thereby bookend the phenomenon to which we refer as `an experiment'.    In this sense, the particle model serves as the \emph{portal} through which the wave model manifests itself---the wave-like behavior of an electron is only indirectly manifested through the pattern generated by localized detections over many runs of the experiment. 
  
\subsection{Persistence \& Nonpersistence}

Wave--particle complementarity suggests that the assumption of \emph{continuous localization,} a notion extrapolated from everyday perception, does not enjoy absolute validity in the microscopic realm.  We propose that the assumption of \emph{persistence} is similarly restricted in its validity.  Furthermore, we propose that the understanding of identical particle-like events requires a synthesis of complementary persistence and nonpersistence models, analogous to the way that localization~(`particle') and delocalization~(`wave') models need to be combined in order to understand individual microscopic entities.

Our proposal is based on the following considerations.  First, through an operational analysis of an experiment in which identical particle-like events are registered at each instant~(Sec.~\ref{sec:persistence}), we have seen that one can construct \emph{two} distinct models.  One of these assumes that successive detections are underpinned by \emph{persistent} underlying entities~(`particles'), whilst the other assumes that no such entities exist, but rather that the events at each instant are the manifestation of a \emph{single} abstract `system'.  

Second, we have demonstrated that these models can be synthesized to derive a quantum symmetrization procedure for describing the behaviour of a system of identical particle-like events.  This derivation not only places a formally \emph{ad hoc} formal procedure on a clear operational and logical foundation, but also naturally resolves the difficulty in reconciling the assertion that identical particles are nonpersistent or not reidentifiable with their manifest reidentifiability in particular situations~(Sec.~\ref{sec:probabilistic-reidentifiability}).

Third, we assert that persistence and nonpersistence are \emph{complementary} descriptions of identical particle-like events on the ground that the persistence and nonpersistence models satisfy the three key features of complementarity, as follows.

First, these two models are \emph{mutually exclusive} in the sense that they make contradictory assumptions about whether or not successive individual detections are underpinned by individual persistent entities.   Consequently, the statement that `this detection was caused by the same object as a previous detection' is meaningful in the first model but not the second.  

Second, despite their contradictory nature, these models can be \emph{synthesized,} with the resulting model capable of describing behaviour that spans the extremes of the original models.  For example, if an electron is sufficiently isolated~(as judged within the persistence model), then, according to the synthetic model, it can be probabilistically reidentified.  But, the synthetic model also describes the behaviour of two electron in a helium atom, a situation in which the persistence model loses its approximate validity.

Third, although synthesized, the original models are not \emph{unified} into a new conceptual scheme.  In particular, the persistence model is needed to calculate the amplitudes that are combined by the symmetrization procedure.  Indeed, without the aid of the persistence model~(wherein one can \emph{analyse} a complex situation into separate persistent components that interact with one another), the calculation of a transition amplitude in the nonpersistence model would appear to be impossible.

\subsection{Parallel between Wave--Particle and Persistence--Nonpersistence Complementarities}
\label{sec:parallel}

The above considerations suggest that there is a close conceptual parallel between wave--particle and the proposed persistence--nonpersistence complementarities.  In order to bring this parallel into sharper focus, we now construct a precise formal parallel between them.

We start by noting that, in Feynman's formulation of quantum theory, one can regard the Feynman amplitude sum rule as a formal encapsulation of Bohr's notion of wave--particle complementarity.  Specifically, in the context of an electron double-slit experiment, there are two models of the electron~(see Table~\ref{tbl:complementarity-comparison}, first column).  In the particle model, the electron is treated as a localized, particle-like entity that traverses one slit or the other in its passage from the source to a given point on the screen.  In the wave model, the electron is treated as a delocalized object that \emph{passes through the silts}.  The amplitude sum rule posits a relationship between the amplitudes in these two models.  

Furthermore, as we have previously shown, the Feynman sum rule can be \emph{derived}, within a suitable operational framework, from a postulate that~(roughly speaking) the amplitudes of the two possible paths in the particle-model determine the amplitude of the `path' in the wave-model~\cite{GKS-PRA}.  In the present context, we interpret this derivation as showing how Bohr's wave--particle complementarity can be viewed as the \emph{basis} of a constructive derivation of the Feynman sum rule, in the same way that we have shown that the proposed persistence--nonpersistence complementarity provides the basis of a derivation of Feynman's symmetrization procedure.

As shown in the Table~\ref{tbl:complementarity-comparison}, this derivation allows us to exhibit a precise formal parallel between the two complementarities.  In each case, two models are synthesized.  One of these models permits an \emph{analysis} of the situation into \emph{parts}---`the electron passes through one slit or the other slit' or `the identical particles make a direct or indirect transition'.  This analysis allows two distinct amplitudes to be defined and, in principle, calculated by making use of the Dirac--Feynman amplitude--action quantization rule~\cite{Goyal2014}.   In contrast, in the other model, no such analysis is possible---all one can say is that `the electron passes through the slits' or that `two identical particle-like events occur at each of two successive times'.  One can associate a transition amplitude with such an unanalyzed process, but its \emph{calculation} appears impossible if one remains within the compass of this model because no corresponding classical model exists.  Although the two models in each case are contradictory in their assumptions, they can be synthesized:~if one posits that the two amplitudes in each analytic model determines the amplitude in the corresponding non-analytic model, it is possible to derive the form of the functional form of the relationship between the amplitudes.

\begin{table*}
\centering
	\begin{ruledtabular}
    \begin{tabular}{p{4cm}p{6cm}p{6cm}} 
    \smallskip
    {\bf } 															& \tableheadtext{Amplitude Sum Rule} 			& \tableheadtext{Symmetrization Procedure} \\ \hline

\smallskip\tableheadtext{Two mutually-incompatible models}   	
																	& \smallskip\tabletext{Particle \& Wave Models}   	
																	& \smallskip\tabletext{Persistence \& Nonpersistence Models} \\ 
    
\medskip\tableheadtext{Synthesis of Models}  
																	&  \vspace{0.18cm}\includegraphics[width=2in]{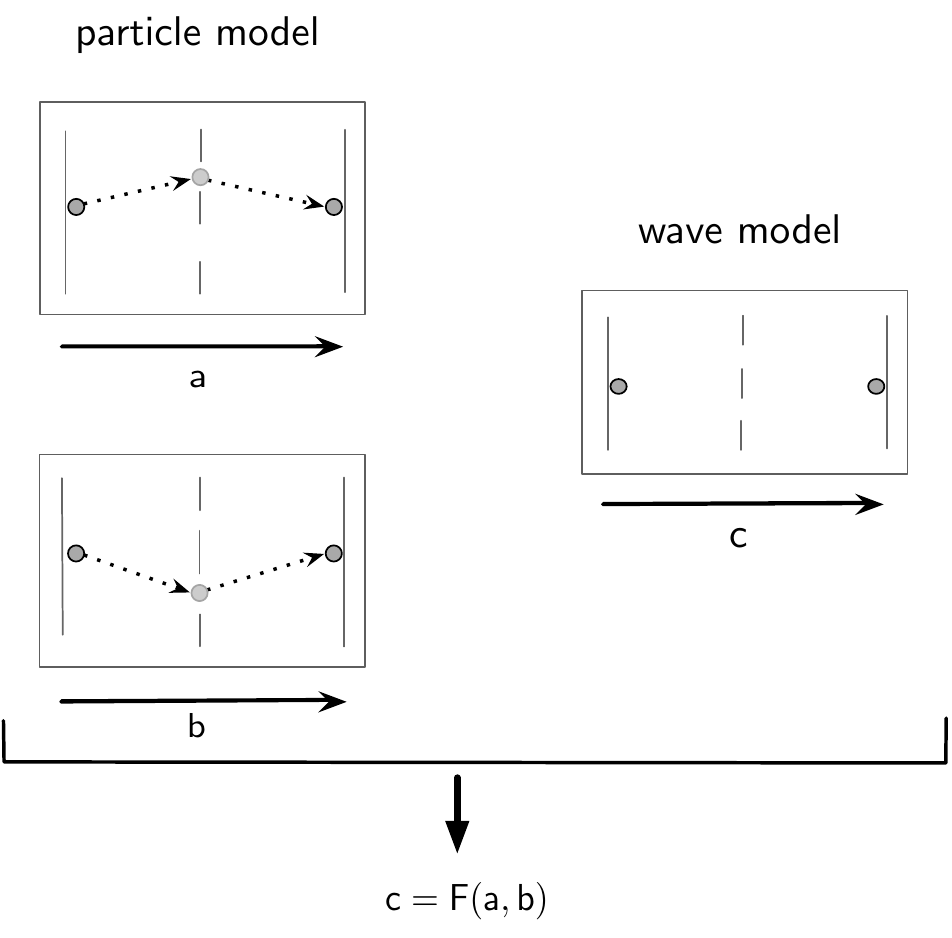}
																	& \vspace{0.16cm}\includegraphics[width=2.5in]{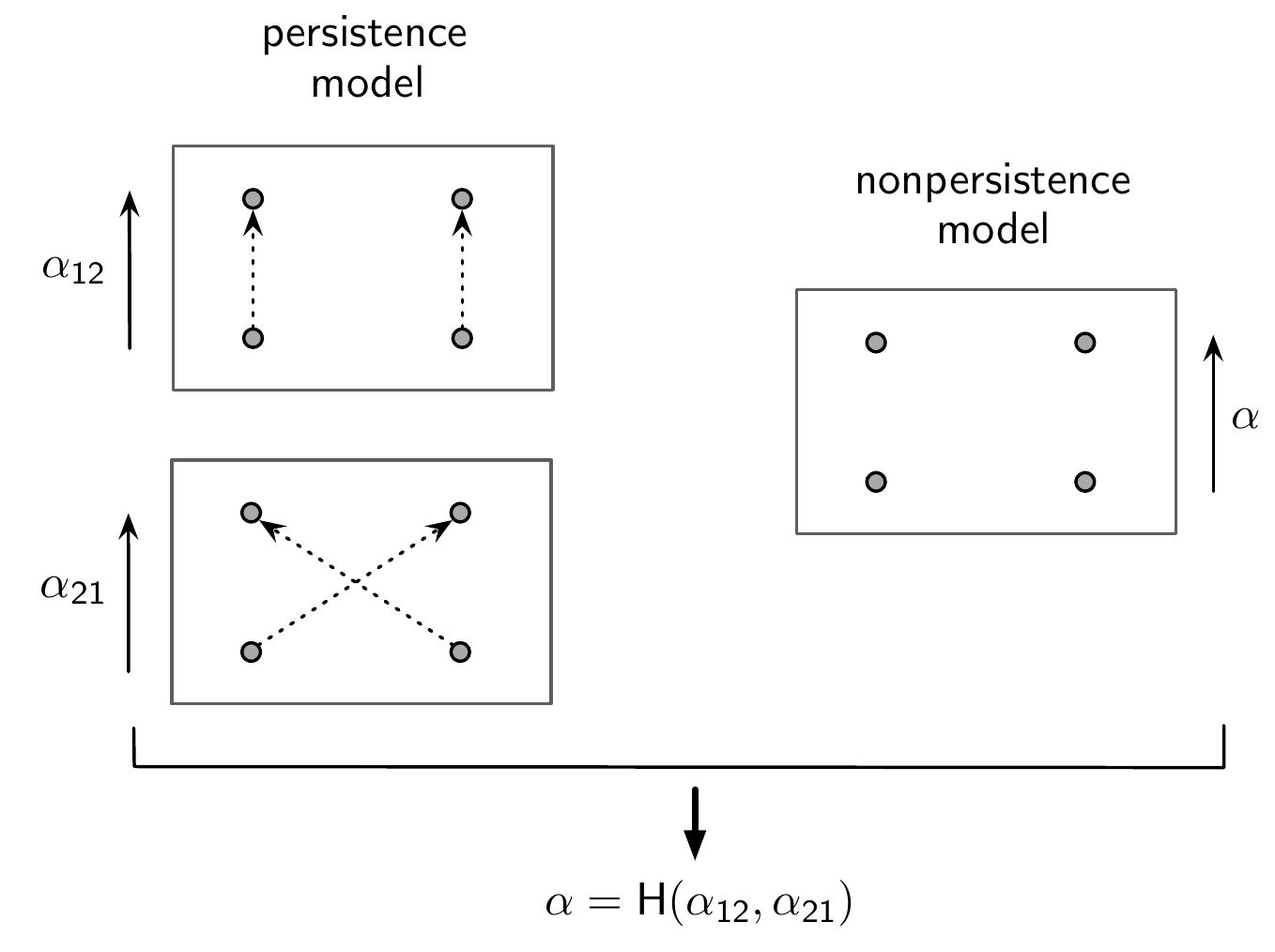} \\ 

\smallskip\tableheadtext{Non-unification}   						
																	& \smallskip\tabletext{1.~Calculation of amplitude~$c$ requires particle model.}
																	& \smallskip\tabletext{1.~Calculation of amplitude~$\alpha$ requires persistence model.}  \\  

																	&  \tabletext{2.~Particle-like detections signify position measurements in particle model, and bookend the phenomenon referred to as `an experiment'.}
																	& \\	
									\\
    \end{tabular}
	\end{ruledtabular}
\caption{\label{tbl:complementarity-comparison} Parallel between the complementarity interpretations of (i)~amplitude sum rule~(exemplified by double slit) and (ii)~symmetrization procedure~(for two identical `particles').  \emph{Case~(i):~}Given detections at a source and screen, the particle model posits that a particle-like object traversed either one slit or the other.  The wave model eschews such an assumption, treating the electron as a delocalized object~(`wave').  These apparently contradictory models can nevertheless be synthesized by positing a functional relationship,~$c = F(a, b)$, between the transitions amplitudes in these models.   As shown in~\cite{GKS-PRA}, the unknown function can be obtained within a broader derivation of Feynman's rules, yielding~$c = a + b$~(Feynman's amplitude sum rule).  Application of the synthetic model requires explicit use of the particle model in order to calculate amplitudes, and use of particle-language to interpret localized detection events.  \emph{Case~(ii):~} The persistence model assumes that detections are underpinned by persistent entities, an assumption eschewed by the nonpersistence model.  Model synthesis is enabled by the assumption that the relation~$\alpha = H(\alpha_{12}, \alpha_{21})$ holds between the amplitudes in these models. Within the Feynman framework,~$H$ can be solved~\cite{Goyal2015} to yield the Feynman form of the symmetrization procedure~$\alpha = \alpha_{12} \pm \alpha_{21}$, where the sign corresponds to fermionic or bosonic behaviour.  Application of this procedure requires use of the persistence model to calculate amplitudes.} 
\end{table*}

An important distinction between the two complementarities, however, deserves to be noted.  As shown in~\cite{GKS-PRA}, the Feynman sum rule can be regarded as arising from a connection between two distinct experimental arrangements---one in which there are which-way detectors at each slit, and the other with a single large detector covering both slits~\cite{GKS-PRA}.  However, the sum rule can \emph{alternatively} be regarded as a connection between two models of the \emph{same} arrangement in which there is a single large detector.  However, in the derivation of the Feynman symmetrization procedure for identical particles given here, only one of these two options is available:~the symmetrization procedure arises from the connection between two distinct models of the \emph{same} experimental arrangement.  Thus, the symmetrization procedure seems to express the notion of complementarity at a deeper level.

\section{Discussion}
\label{sec:discussion}

It is widely believed that identical particles differ from nonidentical particles in that the former lack persistence or reidentifiability.  We have pointed out the deficiencies of such a view, and proposed instead that the specialness of identical particles lies in the fact that \emph{both} persistence and nonpersistence models must be employed in order to cover their full range of behaviour.  We have proved the viability of this viewpoint by showing how the Feynman and Dirac symmetrization procedures that are employed to treat systems of identical particles can be systematically derived through a synthesis of the persistence and nonpersistence models.  We have also indicated how reidentifiability emerges in a context-dependent manner. 

We have further shown that the persistence and nonpersistence models, and the manner of their synthesis, satisfy the key characteristics of Bohr's concept of complementarity.  On this basis, we have proposed that the quantal behaviour of identical particles reflects a \emph{complementarity} of persistence and nonpersistence, analogous to the way in which the behavior of an individual electron is rendered intelligible through Bohr's wave--particle complementarity.  Finally, we have constructed a precise parallel between these two complementarities, which brings their conceptual similarities and dissimilarities into sharper focus.

We conclude with a few brief remarks on which we expect to elaborate elsewhere.

\paragraph{Relationship between the two complementarities.}
The parallel between the persistence--nonpersistence and wave--particle complementarities raises the question of whether there is a single broader perspective from which both complementarities can be seen to emerge, and indeed whether other related complementarities exist.  We leave this question open, apart from noting the presence in each of an essential tension between `whole' and `part'.  That is, in each complementarity, one model permits analysis into `parts'~(either due to the assumption of persistence or the assumption of continuous localization) whereas the other describes the situation as an unanalyzed whole.

\paragraph{Relation between the quantum theoretic and quantum field theoretic models of identical particle-like events.}
According to the thesis put forward here, the symmetrization procedures used in the description of identical particle-like events are a bridge between the quantum-theoretic descriptions of two different models of identical particle-like events.  This bridge allows one to compute evolving states~(or to compute transition amplitudes in the Feynman picture) in the persistence model, and then to combine these in specific ways~(as specified by the Dirac or Feynman symmetrization procedure) to yield evolving states~(or to yield transition amplitudes) in a nonpersistence model. 

As the nonpersistence model regards all the events recorded at each instant as manifestations of an abstract system, the number of these events is a state-determined property~(rather than an intrinsic property) of this system.   Accordingly, a natural generalization of the nonpersistence model considered above would allow a variable number of events to be detected at each instant, and correspondingly allow state-transformations in which the number of events can change.   

One can regard the quantum field theoretic treatment of identical particle-like events as such a generalization.  In the quantum field theoretic model, there no persistent entities~(apart from the abstract system itself), only `excitations'.  The (anti-)~commutation relations amongst the creation and annihilation operators ensure that the states in this model agree with those of the quantum theoretic model for constant event~(or excitation) number.  The formalism allows one to compactly write down superpositions of states with differing numbers of excitations.  The introduction of the notion of persistent individuals, however, requires that one interpret such states within the persistence model---if probabilistic reidentifiability holds in the persistence model~(for each fixed-number component of the superposition), then one can interpret the excitations as persistent `particles'.

\paragraph{Connection to everyday experience.}
We ordinarily assume that the \emph{appearances} perceived in the present moment are underpinned by \emph{objects} that:
\begin{enumerate}[(i)]
\item \emph{Persist} in the time between these appearances; and
\item Assume \emph{forms} that coincide with those of these appearances, not only at the moment of perception but also during the intervening intervals.
\end{enumerate}
As an extrapolation from what is directly perceived, these assumptions constitute a `theory' developed very early in life%
\footnote{As elucidated by Piaget~\cite{Piaget1954}, the concept of object persistence is \emph{learned} by infants, and develops through a series of distinct identifiable stages over the first eighteen or so months of life.  Prior to the development of this concept, infants sometimes behave as if an object removed from sight no longer exists.},  
and are written into the foundation of classical physics.  The above complementarities bring this extrapolation into question, at two distinct levels:
\begin{enumerate}[(i)]
\item \emph{Wave--particle complementarity} brings into question the assumption that an object takes the \emph{same form} between observations as it does during its appearances.  
\item \emph{Persistence--nonpersistence complementarity} brings into question the more basic idea that an object exists between observations and underpins them.  
\end{enumerate}
Nevertheless, the synthesis of complementary models~(wave and particle models; or persistence and nonpersistence models) yields a theory that fits the observations.   Thus, on the one hand, one can regard complementarity as pointing to the limitations of our ordinary models of the appearances; but, on the other, as offering a constructive path to transcend these limitations.



\end{document}